\documentclass[a4paper,11pt]{article}

\usepackage[margin=1in]{geometry}
\usepackage[authoryear]{natbib}
\usepackage[table]{xcolor}
\usepackage{amsmath,amssymb}
\usepackage{graphicx}
\usepackage{booktabs}
\usepackage{hyperref}
\usepackage{xspace}
\usepackage{booktabs}
\usepackage{array}

\usepackage{xurl}
\usepackage{hyperref}
\usepackage{microtype}

\newcolumntype{C}{>{\centering\arraybackslash}p{0.085\textwidth}}
\newcolumntype{L}{>{\raggedright\arraybackslash}p{0.15\textwidth}}

\title{Fuzzy network jump models for soft dynamic clustering of graph-structured data}

\author{
\begin{tabular}{c}
Federico P. Cortese\\
\small University of Milan, Department of Economics, Management, and Quantitative Methods\\
\small National Research Council, Institute for Applied Mathematics and Information Technologies\\
\small 
\href{mailto:federico.cortese@unimi.it}{\texttt{federico.cortese@unimi.it}}
\end{tabular}
}

\date{}

\begin{document}

\maketitle







%

















\begin{abstract}
We introduce a fuzzy network jump model for clustering time-varying observations indexed by the nodes of a weighted graph. The framework allows flexible graph representations with spatial and temporal regularization promoting smooth soft cluster assignments across connected nodes and consecutive time points. Estimation is performed through an efficient alternating optimization scheme that exploits the quadratic structure of the regularization terms. A simulation study covering different levels of spatial dependence and cluster overlap shows that the proposed method accurately recovers the true membership probabilities and  outperforms competing clustering methods. An application to traffic-network data for the city of San Francisco identifies interpretable traffic regimes and reveals their evolution over time and across connected road segments.

\end{abstract}




\maketitle
\noindent\textbf{Keywords:} regime-switching models; soft clustering; space-time interaction; transport modeling; unsupervised learning

\section{Introduction}
\label{sec:intro}

Clustering of graph nodes aims to identify groups of vertices that are connected, structurally similar, or characterized by comparable node-level attributes
\citep{schaeffer2007graph}.
An additional layer of complexity arises when the clustering structure changes over time, giving rise to a \textit{dynamic} clustering problem.
Relevant examples include social and biological networks, as well as transport systems. In the latter, dynamic node clustering can support the identification of spatially coherent traffic regimes and recurrent congestion patterns \citep{saeedmanesh2017dynamic}, possibly assisting transport planners in targeting interventions, evaluating network performance, and improving infrastructure management.
In these applications, nodes may represent traffic sensors, mobility zones, or public transport stations, whereas edges may encode geographic proximity, traffic flows, travel times, or other measures of interaction intensity. Transport networks may therefore differ substantially in the meaning and scale of their edge weights.
Moreover, node-level observations are often multivariate and heterogeneous, requiring a clustering framework flexible enough to accommodate different data types while capturing temporal dynamics.

Graph clustering methods can be broadly grouped into three main families. Topology-based approaches, including spectral clustering, graph-cut methods, and modularity optimization, partition the nodes by exploiting the structure of a weighted adjacency or affinity matrix
\citep{vonluxburg2007tutorial,fortunato2010community}. These methods are computationally attractive and require relatively few distributional assumptions, but they typically produce hard partitions, and incorporate node attributes and temporal dependence only indirectly. 
%
A second family is represented by probabilistic network models, particularly stochastic block models and their weighted, mixed-membership, and dynamic extensions \citep{nowicki2001estimation,aicher2015learning, palowitch2018significance,matias2017statistical,lee2019review}.  These models provide an explicit probabilistic description of the network, but require the distribution of the edge weights to be specified in advance. 
They also assume that nodes assigned to the same block have similar connectivity patterns, an assumption that may be restrictive in heterogeneous transport networks. Dynamic extensions further increase computational complexity, and often require explicit assumptions on the temporal evolution of block memberships or interaction parameters.
A third family includes neural network-based approaches,
which learn low-dimensional node representations from network structure, node attributes, and temporal information \citep{kipf2016variational,hamilton2017representation, singer2019node2bits,barros2021survey}. Although flexible and suitable for complex data, they are often sensitive to tuning choices and difficult to interpret.

The present work adopts a different perspective. Rather than treating the edge weights as realizations from a prescribed network-generating distribution or using the graph solely to construct a latent embedding, the graph is introduced as a general regularization structure within a clustering model for multivariate node-level observations. 
Edge weights determine the strength with which connected nodes are encouraged to exhibit similar cluster memberships and can be flexibly defined in terms of spatial adjacency, geographic distance, network connectivity, or any other application-specific measure of association.

The proposed approach, which we refer to as \textit{fuzzy network jump model} (FNJM), allows each node to exhibit partial membership in multiple regimes.
A temporal regularization component further promotes persistence of the memberships over time while preserving the possibility of detecting regime changes, whereas a spatial regularization component encourages similar memberships across connected nodes according to the strength of their connection.
The resulting framework lies between purely topological community-detection methods and fully generative network models. It retains direct interpretability in terms of the original node-level variables, accommodates heterogeneous graph structures, and provides a flexible tool for the dynamic analysis of graph-structured data.

The proposed framework builds on statistical jump models, a class of unsupervised learning methods introduced by \citet{bemporad:2018} as an alternative to hidden Markov models \citep{bart:farc:penn:13,zucchini:2017} for dynamic modeling of multivariate time series data. Jump models essentially describe the temporal evolution of a complex system through transitions among latent regimes, each represented by a simpler local model inferred from the observed data. By avoiding a fully specified probabilistic state-space structure, they offer a flexible framework for capturing persistent regimes and abrupt changes in complex dynamics.
Recent extensions incorporate sparse \citep{nystrup:2021} and state-conditional \citep{cortese2026robust} variable selection, soft clustering of time series data \citep{aydinhan2024identifying,cortese2025fuzzy}, and spatio-temporal hard clustering \citep{cortese2026spatio}.

A simulation study based on data generated under different spatial-dependence and fuzziness scenarios shows that the proposed method accurately recovers the true membership probabilities and outperforms competing models in terms of both soft-membership estimation and cluster-assignment recovery, while an application to the San Francisco traffic network illustrates its ability to identify interpretable congestion states and track their evolution over time and across connected road segments.

The rest of the paper is organized as follows. Section~\ref{sec:meth} introduces the proposed method and the estimation procedure. Section~\ref{sec:simstud} presents the simulation study, while Section~\ref{sec:app} describes the empirical application. Section~\ref{sec:disc} concludes the paper. Appendix \ref{app:add_simstud} presents additional simulation results.

\section{Methodology}
\label{sec:meth}

Let $\boldsymbol{z}_{t,m}=(z_{tm1},\ldots,z_{tmP})^\prime$
denote the $P$-dimensional vector of observations collected at time $t=1,\ldots,T$
and location $m=1,\ldots,M$. 
We assume that each observation is associated with a membership vector
\[
\boldsymbol{u}_{t,m}
=
(u_{tm1},\ldots,u_{tmK})^\prime
\in \Delta_{K-1},
\]
where
\[
\Delta_{K-1}
=
\left\{
\boldsymbol{u}\in[0,1]^K:
\sum_{k=1}^K u_k=1
\right\},
\]
is the probability simplex of dimension $K-1$. The collection of all
membership vectors is denoted by $\boldsymbol{U}$, while
$\boldsymbol{\mu}_k$ denotes the $P$-dimensional prototype vector associated with cluster
$k$.

\subsection{Spatial graph and weights}

The spatial structure is represented by an undirected weighted graph
$G=(V,E)$, where $V=\{1,\ldots,M\}$ is the set of locations and
$E$ is the set of edges connecting pairs of distinct locations. Let
\[
\widetilde{\boldsymbol{W}}
=
(\widetilde{w}_{im})_{i,m=1}^M,
\]
be the corresponding symmetric non-negative weight matrix, satisfying
\[
\widetilde{w}_{im}=\widetilde{w}_{mi}\geq 0,
\qquad
\widetilde{w}_{mm}=0,
\qquad
\widetilde{w}_{im}>0
\iff
(i,m)\in E.
\]

The graph can be constructed from an adjacency matrix, an edge list, or a
symmetric dissimilarity matrix. In the latter case, given a threshold $h>0$,
an edge is retained whenever
\(
d_{im}\leq h,
\)
and its raw weight can be defined through a non-increasing kernel
$\kappa(\cdot)$ as
\(
\widetilde{w}_{im}
=
\kappa(d_{im})\mathbb{I}(d_{im}\leq h).
\)
The dissimilarity $d_{im}$ 
may represent geodesic
distance, network distance, travel time, or another application-specific
notion of spatial proximity.
Optionally, the raw weight matrix can be symmetrically degree-normalized. Let
\[
d_i=\sum_{m=1}^M \widetilde{w}_{im},
\]
then define
\[
\boldsymbol{W}
=
\boldsymbol{D}^{-1/2}
\widetilde{\boldsymbol{W}}
\boldsymbol{D}^{-1/2},
\qquad
\boldsymbol{D}
=
\operatorname{diag}(d_1,\ldots,d_M).
\]
This normalization prevents highly connected locations from dominating the spatial penalty and ensures a more balanced regularization across nodes with different degrees.
Once the graph has been constructed, the estimator stores only the edge list
$\{(i,m,w_{im}):(i,m)\in E\}$. Consequently, spatial computations are
performed over the observed edges rather than over all
$\binom{M}{2}$ pairs of locations.


The model uses the \citet{gower1971general} dissimilarity between observation $\boldsymbol{z}_{t,m}$ and prototype $\boldsymbol{\mu}_k$, defined as
\[
g(\boldsymbol{z}_{t,m},\boldsymbol{\mu}_k)
=
\frac{1}{P}
\sum_{p=1}^P
\delta_p(z_{tmp},\mu_{kp}),
\]
where, for a continuous feature,
\[
\delta_p(z_{tmp},\mu_{kp})
=
\frac{|z_{tmp}-\mu_{kp}|}{R_p},
\]
with $R_p$ denoting the observed range of feature $p$, whereas for a categorical feature,
\[
\delta_p(z_{tmp},\mu_{kp})
=
\mathbb{I}(z_{tmp}\neq\mu_{kp}).
\]
%

\noindent
Finally, the proposed estimator minimizes
\begin{equation}
\label{eq:STF}
\begin{aligned}
F(\boldsymbol{U},\boldsymbol{\mu})
={}&
\sum_{t=1}^{T}
\sum_{m=1}^{M}
\sum_{k=1}^{K}
u_{tmk}^{\,\phi}
g(\boldsymbol{z}_{t,m},\boldsymbol{\mu}_k)
\\
&+
\frac{\lambda_T}{2}
\sum_{t=2}^{T}
\sum_{m=1}^{M}
\left\|
\boldsymbol{u}_{t,m}
-
\boldsymbol{u}_{t-1,m}
\right\|_2^2
\\
&+
\frac{\lambda_S}{2}
\sum_{t=1}^{T}
\sum_{(i,m)\in E}
w_{im}
\left\|
\boldsymbol{u}_{t,i}
-
\boldsymbol{u}_{t,m}
\right\|_2^2,
\end{aligned}
\end{equation}
subject to
\(
\boldsymbol{u}_{t,m}\in\Delta_{K-1},
\,
t=1,\ldots,T,\, m=1,\ldots,M.
\)

The parameter $\phi \geq 1$ controls the degree of fuzziness in the data-fitting term. When $\lambda_T=\lambda_S=0$ and $\phi=1$, the model reduces to a Gower-based version of the $k$-prototypes algorithm \citep{huang1998extensions}. For $\phi>1$, fuzziness\footnote{Fractional memberships may also arise when $\phi=1$ if either the temporal or spatial regularization parameter is strictly positive. This occurs because the regularization terms penalize abrupt differences between neighbouring membership vectors, so an interior simplex solution may provide a better compromise between data fit and spatio-temporal smoothness even when $\phi=1$.} is directly induced by the membership exponent \citep{dunn1973fuzzy,bezdek1981pattern}.
For one-hot memberships,
\(
\frac{1}{2}\|\boldsymbol{u}-\boldsymbol{v}\|_2^2
=
\mathbb{I}(c_u\neq c_v),
\)
where $c_u$ and $c_v$ denote the corresponding cluster labels. Hence, in the hard-clustering limit, the spatial regularization term reduces to a weighted Potts disagreement penalty \citep{wu1982potts}.

The parameters $\lambda_T,\lambda_S\geq0$ control temporal and spatial regularization, with larger values promoting smoother memberships over time and across connected locations, respectively.
To further characterize the spatial regularization term, let
\[
\boldsymbol{D}_W
=
\operatorname{diag}(d_1^W,\ldots,d_M^W),
\qquad
d_i^W=\sum_{m=1}^M w_{im},
\]
and define the weighted graph Laplacian as
\(
\boldsymbol{L}_W
=
\boldsymbol{D}_W-\boldsymbol{W}.
\)
Moreover, let \(\boldsymbol{U}_t\) be the \(M\times K\) matrix whose \(m\)-th row is
\(\boldsymbol{u}_{t,m}^{\prime}\), then the spatial regularization term in Eq. \eqref{eq:STF} can be equivalently written as
\[
\frac{\lambda_S}{2}
\sum_{t=1}^T
\operatorname{tr}
\left(
\boldsymbol{U}_t^\prime
\boldsymbol{L}_W
\boldsymbol{U}_t
\right),
\]
which makes explicit that the penalty smooths membership vectors over the graph, with the contribution of each connection determined by its weight.
%
%
%

In Equation \ref{eq:STF}, we adopt the squared $\ell_2$-penalty for the spatial and temporal components because it is smooth, has a simple linear gradient, and admits a convenient graph Laplacian representation. 
Additionally, the quadratic penalty further discourages large differences between neighbouring membership vectors, while the factor \(1/2\) ensures consistency with the statistical jump model of \citet{nystrup:2020} in the hard-clustering limit.
An $\ell_1$ alternative, similar to that of \citet{aydinhan2024identifying}, can also be used, but it is computationally less convenient because its non-smoothness complicates gradient-based optimization.

\subsection{Estimation}

The objective in \eqref{eq:STF} is minimized by block coordinate descent,
alternating between membership and prototype updates. Since the complete
problem is non-convex, the algorithm is combined with multiple initializations.
Specifically, each initialization uses a Gower-based analogue of the $k$-means++ strategy \citep{arthur2007k}.

\begin{itemize}
    \item[] \textit{\textbf{Membership update}}.
%
For fixed prototypes, the membership vectors remain coupled through the
spatial and temporal penalties. 
The algorithm performs a Gauss-Seidel sweep \citep{grippo2000convergence, tseng2001convergence} over the
$T\times M$ time-location pairs.
Conditionally on the current values of all neighbouring memberships, the local problem for $\boldsymbol{u}_{t,m}$ is
\begin{equation}
\label{eq:local-membership}
\min_{\boldsymbol{u}\in\Delta_{K-1}}
\,\,
h_{tm}(\boldsymbol{u})
=
\sum_{k=1}^K
u_k^\phi
g(\boldsymbol{z}_{t,m},\boldsymbol{\mu}_k)
+
\frac{\lambda_T}{2}
\sum_{r\in\mathcal{B}_T(t)}
\left\|
\boldsymbol{u}
-
\boldsymbol{u}_{r,m}
\right\|_2^2
+
\frac{\lambda_S}{2}
\sum_{i\in\mathcal{B}_S(m)}
w_{im}
\left\|
\boldsymbol{u}
-
\boldsymbol{u}_{t,i}
\right\|_2^2,
\end{equation}
where $\mathcal{B}_T(t)$ contains the available previous and next time points,
and $\mathcal{B}_S(m)$ is the set of spatial neighbours of location $m$.
%
The $k$-th component of the local
gradient is
\[
\begin{aligned}
\frac{\partial h_{tm}}{\partial u_k}
=
\phi\,
g(\boldsymbol{z}_{t,m},\boldsymbol{\mu}_k)
u_k^{\phi-1}
+
\lambda_T
\sum_{r\in\mathcal{B}_T(t)}
\left(
u_k-u_{rmk}
\right)
+
\lambda_S
\sum_{i\in\mathcal{B}_S(m)}
w_{im}
\left(
u_k-u_{tik}
\right),
\end{aligned}
\]
which is solved by projected gradient descent \citep{duchi2008efficient}
\[
\boldsymbol{u}^{(\ell+1)}
=
\Pi_{\Delta_{K-1}}
\left[
\boldsymbol{u}^{(\ell)}
-
\eta_\ell
\nabla h_{tm}(\boldsymbol{u}^{(\ell)})
\right],
\]
where $\Pi_{\Delta_{K-1}}$ denotes the Euclidean projection onto the probability
simplex. A backtracking line search reduces $\eta_\ell$ until the local
objective does not increase. 
The Gauss-Seidel updates alternate between forward and reverse order across 
iterations, reducing sensitivity to the visiting order.

\item[] \textit{\textbf{Prototype update}}.
%
For fixed memberships, the prototype of cluster $k$ solves
\begin{equation}
\label{eq:proto_update}
\boldsymbol{\mu}_k
=
\arg\min_{\boldsymbol{\mu}}
\sum_{t=1}^T
\sum_{m=1}^M
u_{tmk}^{\,\phi}
g(\boldsymbol{z}_{t,m},\boldsymbol{\mu}).
\end{equation}
Because the Gower dissimilarity is additive across features, the update can be
performed feature by feature. \citet{cortese2025fuzzy} show that the minimizers of \eqref{eq:proto_update} are the
weighted medians for continuous features, and the weighted modes for categorical ones, with weights given by $u_{tmk}^{\,\phi}$.

\end{itemize}

\noindent
After updating the prototypes, all observation-prototype Gower
dissimilarities are recomputed. Membership and prototype updates are repeated
until
\[
\left|
F^{(r-1)}-F^{(r)}
\right|
\leq
\varepsilon
\left(
1+\left|F^{(r-1)}\right|
\right),
\]
or until a prescribed maximum number of 
iterations is reached. The
backtracking membership updates and exact feature-wise prototype updates yield
a non-increasing objective up to numerical tolerance.
The computational complexity of model estimation is approximately $\mathcal{O}\!\left(KPTM\log(TM)\right)$. On a MacBook Pro equipped with a 10-core Apple M5 processor and 16\,GB of RAM, estimation of the final FNJM model with eight initializations required on average \(4.66\) seconds.
Code for reproducibility, including a toy example and the empirical analysis of Section~\ref{sec:app}, is available at \url{https://github.com/FedericoCortese/FNJM}.

\section{Simulation study}
\label{sec:simstud}

We conduct a simulation study to evaluate the ability of the proposed model to
recover a time-varying distribution of fuzzy memberships under different
degrees of spatial dependence and cluster overlap. 
The data-generating process
combines spatially correlated latent score fields with temporally persistent dynamics and heavy-tailed observations.
We independently sample \(M\) locations from the uniform distribution on \([0,1]^2\), and let \(d_{im}\) denote the Euclidean distance between locations \(i\) and \(m\). The estimation
graph is constructed using a threshold \(h\), with raw weights
\(
\widetilde w_{im}
=
\exp(-d_{im}/h)\mathbb{I}(d_{im}\leq h).
\)
Then, the resulting weights are symmetrically degree-normalized before model
estimation. 
Following the spatial construction of \citet{paci:2018}, we define the
dense spatial correlation matrix
\[
\boldsymbol{\Gamma}_{\alpha}
=
\left\{
\exp(-\alpha d_{im})
\right\}_{i,m=1}^{M},
\]
where \(\alpha>0\) controls the decay of spatial dependence. 
%
For each cluster \(k=1,\ldots,K\), we independently generate a
spatio-temporal latent score field
\(\boldsymbol{a}_{t,k}=(a_{t1k},\ldots,a_{tMk})^\prime\) according to
\begin{align}
\boldsymbol{a}_{1,k}
&\sim
\mathcal N_M
\left(
\mathbf 0,
\tau^2\boldsymbol{\Gamma}_{\alpha}
\right),
\\
\boldsymbol{a}_{t,k}
&=
\beta\boldsymbol{a}_{t-1,k}
+
\tau\sqrt{1-\beta^2}\,
\boldsymbol{\varepsilon}_{t,k},
\qquad
\boldsymbol{\varepsilon}_{t,k}
\sim
\mathcal N_M
\left(
\mathbf 0,
\boldsymbol{\Gamma}_{\alpha}
\right),
\quad t=2,\ldots,T.
\label{eq:latent-score-process}
\end{align}
The factor \(\sqrt{1-\beta^2}\) ensures that \(\tau^2\) is the stationary
marginal variance of each latent score, independently of the value of
\(\beta\). We fix \(\beta=0.9\), corresponding to strong temporal
persistence.
The true fuzzy memberships are then obtained through the location- and
time-specific softmax transformation
\begin{equation}
u^\star_{tmk}
=
\frac{\exp(a_{tmk})}
{\sum_{h=1}^{K}\exp(a_{tmh})},
\qquad
k=1,\ldots,K,
\label{eq:true-soft-memberships}
\end{equation}
so that \(u^\star_{tmk}\in(0,1)\) and
\(\sum_{k=1}^{K}u^\star_{tmk}=1\). 
%
For each space-time observation, a latent component is sampled according to
\[
s_{tm}\mid\boldsymbol{u}^\star_{t,m}
\sim
\operatorname{Categorical}
\left(
u^\star_{tm1},\ldots,u^\star_{tmK}
\right).
\]
Conditional on \(s_{tm}=k\), the \(P\)-dimensional observation is generated as
\begin{equation}
\boldsymbol{z}_{t,m}\mid s_{tm}=k
\sim
\mathcal T_{P,\nu}
\left(
\boldsymbol{\mu}_k,
\frac{\nu-2}{\nu}\mathbf I_P
\right),
\label{eq:simulation-observation-model}
\end{equation}
where \(\mathcal T_{P,\nu}(\boldsymbol{\mu},\boldsymbol{\Sigma})\) denotes a
multivariate Student-\(t\) distribution with location
\(\boldsymbol{\mu}\), scale matrix \(\boldsymbol{\Sigma}\), and \(\nu\)
degrees of freedom. 

We set \(K=2\), with component locations
\(
\boldsymbol{\mu}_1=(-1,-1,-1)^\prime,
\,
\boldsymbol{\mu}_2=(1,1,1)^\prime.
\)
The number of features is fixed at \(P=3\), which, together with the chosen values of \(T=200\) and \(M=300\), essentially mimic the traffic application of Section \ref{sec:app}.
Finally, we set the degrees of freedom to \(\nu=4\).

We vary the spatial correlation decay parameter as $\alpha\in\{0.01,1\}$, corresponding to strong (S-) and weak (W-) spatial dependence, respectively. We also vary the latent-score scale as $\tau\in\{0.2,5\}$, yielding soft (-S) and nearly hard (-H) memberships, respectively. This choice of parameters results in the four scenarios reported in Table~\ref{tbl:scenarios}. For each scenario, 100 independent datasets are generated using different random seeds.
\begin{table}
\small
\caption{Simulation scenarios considered in the study.}
\label{tbl:scenarios}
\begin{tabular*}{\textwidth}{@{\extracolsep{\fill}}cclc@{}}
\toprule
$\alpha$ & $\tau$ & Scenario & Scenario label\\
\midrule
0.01 & 0.2 & Strong spatial dependence, soft memberships&S-S \\
0.01 & 5.0 & Strong spatial dependence, hard memberships&S-H \\
1.00 & 0.2 & Weak spatial dependence, soft memberships&W-S \\
1.00 & 5.0 & Weak spatial dependence, hard memberships&W-H \\
\bottomrule
\end{tabular*}
\end{table}



We fix the distance threshold at \(h=0.05\) for all scenarios and
replications. For \(M=300\) locations independently sampled from the unit
square, the threshold graph has an average density of approximately \(0.76\%\)
and an average degree of \(2.26\), closely matching the density of \(0.72\%\)
and average degree of \(2.21\) observed in the traffic application. Since the
threshold graph may be disconnected, we augment its edge set with the
Euclidean minimum spanning tree to ensure connectivity.
Appendix \ref{app:add_simstud} presents a complementary simulation study based on a substantially denser estimation graph.

\subsection{Model fitting and performance assessment}

For each simulated dataset, FNJM is fitted over the grid
\[
\lambda_T,\lambda_S
\in
\{0,0.1,\ldots,1\},
\,
\phi
\in
\{1,1.25,1.50,1.75,2\},
\]
which contains
\(11\times11\times5=605\) configurations for each simulated dataset.

We compare the proposed method with four competitors: the \(k\)-prototypes
algorithm of \citet{huang1998extensions}, the statistical jump model of
\citet{nystrup:2020}, the fuzzy jump model of \citet{cortese2025fuzzy}, and a Gaussian hidden Markov random-field model \citep[G-HMRF,][]{yang2022scmeb}
that incorporates spatial dependence through a neighbourhood graph and is
used here as a purely spatial benchmark.
For each method and scenario, the reported configuration of hyperparameters is the one minimizing the median membership root mean squared error (RMSE); balanced accuracy (BAC) and the adjusted Rand index \citep[ARI,][]{hubert:1985} are used as secondary criteria in the event of ties. 
Each fit uses one initialization, 10 iterations, and a numerical tolerance of \(10^{-4}\).
To overcome label switching, the estimated memberships are first aligned with the true memberships by selecting the permutation of cluster labels that minimizes the total squared membership error.

Let
\(\widehat u_{tmk}\) denote the aligned estimated membership. 
The performance of the competing models in recovering the true fuzzy memberships is compared using the RMSE and cross-entropy (CE), defined as
\begin{align}
\operatorname{RMSE}
&=
\left[
\frac{1}{TMK}
\sum_{t=1}^{T}
\sum_{m=1}^{M}
\sum_{k=1}^{K}
\left(
\widehat u_{tmk}-u^\star_{tmk}
\right)^2
\right]^{1/2},
\\
\operatorname{CE}
&=
-\frac{1}{TM}
\sum_{t=1}^{T}
\sum_{m=1}^{M}
\sum_{k=1}^{K}
u^\star_{tmk}
\log\!\left(\widehat u_{tmk}\right).
\end{align}
%
To evaluate recovery of the dominant regimes, we additionally compare the \textit{maximum-a-posteriori} regime classification
\(
\widehat s_{tm}
=
\arg\max_k\widehat u_{tmk},
\)
with its ground truth counterpart
\(
s^\star_{tm}
=
\arg\max_k u^\star_{tmk}
\), 
using BAC and ARI. 

\subsection{Simulation results and sensitivity analysis}

\begin{figure}
    \centering
    \includegraphics[width=1\linewidth]{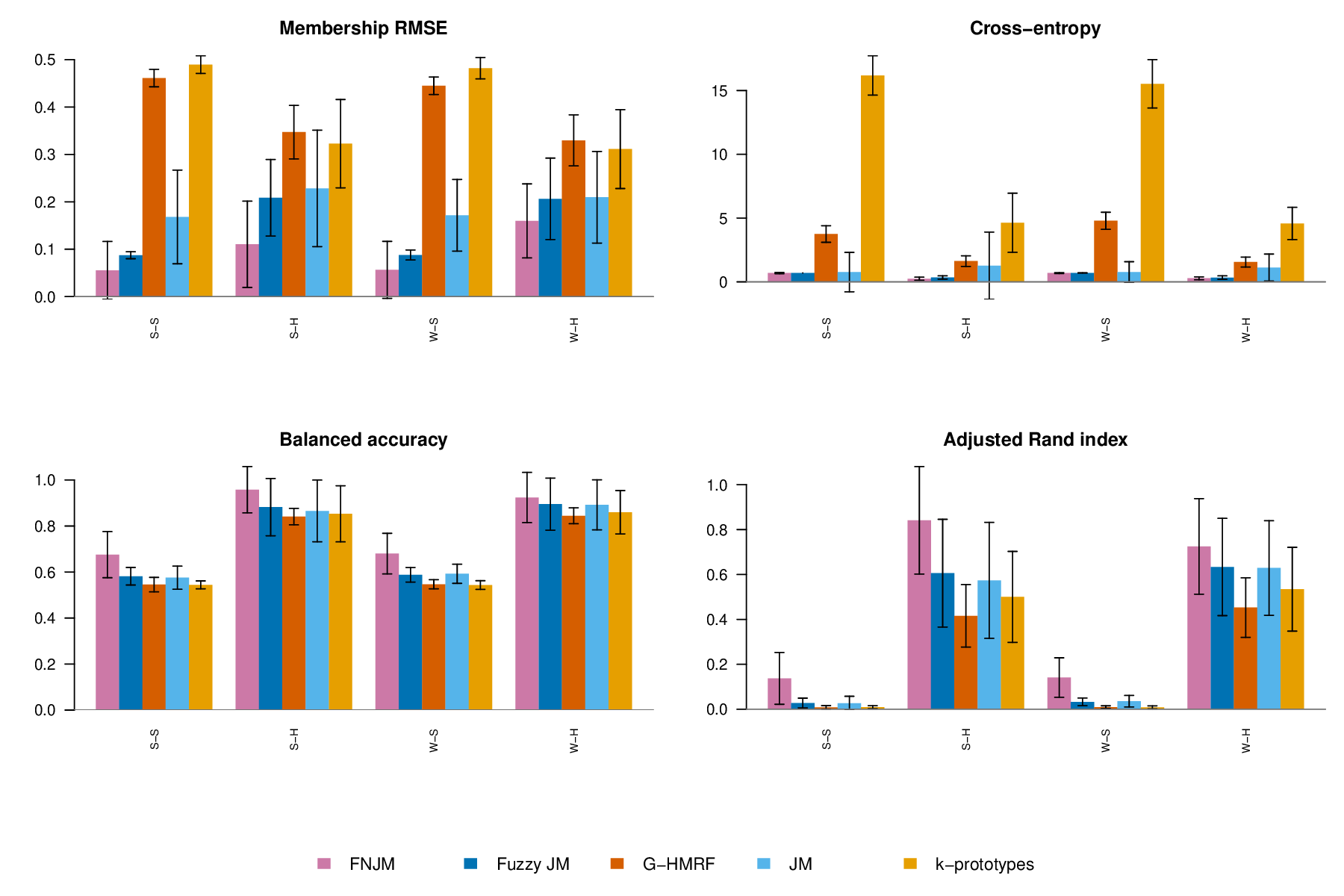}
    \caption{Performance comparison across simulation scenarios based on membership RMSE, cross-entropy, balanced accuracy, and the adjusted Rand index. Bars show median values over 100 replicates, with standard deviations as error bars.}
    \label{fig:model-performance}
\end{figure}

Table~\ref{tbl:simulation-results} reports the median RMSE, CE, BAC, and ARI over 100 replicates for each method and scenario, with standard deviations in parentheses. FNJM achieves the best performance across all scenarios, attaining the lowest RMSE and CE and the highest BAC and ARI. Its advantage is particularly pronounced under strong spatial dependence, where it reaches a BAC of \(0.958\) and an ARI of \(0.841\) in the S-H scenario. Among the competing methods, 
Fuzzy JM generally performs best, while the relative performance of JM, G-HMRF, and \(k\)-prototypes varies across scenarios and metrics.
The S-S and W-S scenarios are the most challenging in terms of classification performance, with FNJM attaining an ARI of \(0.137\) and 0.141, respectively. This is expected because soft memberships make the latent states less clearly separated, even when spatial dependence is strong. In fact, RMSEs remain satisfactory in both cases.
\begin{table}
\caption{Simulation results for each method across scenarios. Entries report median performance over 100 replicates, with standard deviations in parentheses. For each scenario, the lowest median RMSE and CE and the highest median BAC and ARI are shown in bold.}
\label{tbl:simulation-results}
\begin{tabular*}{\textwidth}{@{\extracolsep{\fill}}CCLCCCC@{}}
\toprule
$\alpha$ & $\tau$ & Method & RMSE & CE & BAC & ARI
\tabularnewline
\midrule

\multicolumn{7}{c}{\textbf{S-S}}
\tabularnewline
\midrule
0.01 & 0.2 & FNJM
& \shortstack[c]{\textbf{0.055}\tabularnewline{\scriptsize (0.061)}}
& \shortstack[c]{\textbf{0.691}\tabularnewline{\scriptsize (0.045)}}
& \shortstack[c]{\textbf{0.676}\tabularnewline{\scriptsize (0.100)}}
& \shortstack[c]{\textbf{0.137}\tabularnewline{\scriptsize (0.115)}}
\tabularnewline
& & Fuzzy JM
& \shortstack[c]{0.087\tabularnewline{\scriptsize (0.007)}}
& \shortstack[c]{0.700\tabularnewline{\scriptsize (0.003)}}
& \shortstack[c]{0.582\tabularnewline{\scriptsize (0.038)}}
& \shortstack[c]{0.028\tabularnewline{\scriptsize (0.022)}}
\tabularnewline
& & G-HMRF
& \shortstack[c]{0.461\tabularnewline{\scriptsize (0.018)}}
& \shortstack[c]{3.753\tabularnewline{\scriptsize (0.650)}}
& \shortstack[c]{0.545\tabularnewline{\scriptsize (0.032)}}
& \shortstack[c]{0.009\tabularnewline{\scriptsize (0.008)}}
\tabularnewline
& & JM
& \shortstack[c]{0.168\tabularnewline{\scriptsize (0.099)}}
& \shortstack[c]{0.766\tabularnewline{\scriptsize (1.546)}}
& \shortstack[c]{0.575\tabularnewline{\scriptsize (0.050)}}
& \shortstack[c]{0.027\tabularnewline{\scriptsize (0.031)}}
\tabularnewline
& & $k$-prototypes
& \shortstack[c]{0.490\tabularnewline{\scriptsize (0.019)}}
& \shortstack[c]{16.183\tabularnewline{\scriptsize (1.542)}}
& \shortstack[c]{0.544\tabularnewline{\scriptsize (0.017)}}
& \shortstack[c]{0.009\tabularnewline{\scriptsize (0.007)}}
\tabularnewline

\midrule
\multicolumn{7}{c}{\textbf{S-H}}
\tabularnewline
\midrule
0.01 & 5.0 & FNJM
& \shortstack[c]{\textbf{0.110}\tabularnewline{\scriptsize (0.091)}}
& \shortstack[c]{\textbf{0.248}\tabularnewline{\scriptsize (0.120)}}
& \shortstack[c]{\textbf{0.958}\tabularnewline{\scriptsize (0.101)}}
& \shortstack[c]{\textbf{0.841}\tabularnewline{\scriptsize (0.239)}}
\tabularnewline
& & Fuzzy JM
& \shortstack[c]{0.209\tabularnewline{\scriptsize (0.081)}}
& \shortstack[c]{0.345\tabularnewline{\scriptsize (0.128)}}
& \shortstack[c]{0.882\tabularnewline{\scriptsize (0.124)}}
& \shortstack[c]{0.606\tabularnewline{\scriptsize (0.240)}}
\tabularnewline
& & G-HMRF
& \shortstack[c]{0.347\tabularnewline{\scriptsize (0.056)}}
& \shortstack[c]{1.628\tabularnewline{\scriptsize (0.416)}}
& \shortstack[c]{0.841\tabularnewline{\scriptsize (0.036)}}
& \shortstack[c]{0.416\tabularnewline{\scriptsize (0.139)}}
\tabularnewline
& & JM
& \shortstack[c]{0.228\tabularnewline{\scriptsize (0.123)}}
& \shortstack[c]{1.270\tabularnewline{\scriptsize (2.634)}}
& \shortstack[c]{0.866\tabularnewline{\scriptsize (0.134)}}
& \shortstack[c]{0.574\tabularnewline{\scriptsize (0.258)}}
\tabularnewline
& & $k$-prototypes
& \shortstack[c]{0.323\tabularnewline{\scriptsize (0.093)}}
& \shortstack[c]{4.633\tabularnewline{\scriptsize (2.317)}}
& \shortstack[c]{0.853\tabularnewline{\scriptsize (0.122)}}
& \shortstack[c]{0.500\tabularnewline{\scriptsize (0.203)}}
\tabularnewline

\midrule
\multicolumn{7}{c}{\textbf{W-S}}
\tabularnewline
\midrule
1.00 & 0.2 & FNJM
& \shortstack[c]{\textbf{0.056}\tabularnewline{\scriptsize (0.060)}}
& \shortstack[c]{\textbf{0.691}\tabularnewline{\scriptsize (0.039)}}
& \shortstack[c]{\textbf{0.680}\tabularnewline{\scriptsize (0.088)}}
& \shortstack[c]{\textbf{0.141}\tabularnewline{\scriptsize (0.088)}}
\tabularnewline
& & Fuzzy JM
& \shortstack[c]{0.088\tabularnewline{\scriptsize (0.011)}}
& \shortstack[c]{0.699\tabularnewline{\scriptsize (0.005)}}
& \shortstack[c]{0.588\tabularnewline{\scriptsize (0.032)}}
& \shortstack[c]{0.033\tabularnewline{\scriptsize (0.017)}}
\tabularnewline
& & G-HMRF
& \shortstack[c]{0.445\tabularnewline{\scriptsize (0.019)}}
& \shortstack[c]{4.788\tabularnewline{\scriptsize (0.670)}}
& \shortstack[c]{0.547\tabularnewline{\scriptsize (0.020)}}
& \shortstack[c]{0.009\tabularnewline{\scriptsize (0.006)}}
\tabularnewline
& & JM
& \shortstack[c]{0.172\tabularnewline{\scriptsize (0.076)}}
& \shortstack[c]{0.771\tabularnewline{\scriptsize (0.811)}}
& \shortstack[c]{0.592\tabularnewline{\scriptsize (0.042)}}
& \shortstack[c]{0.036\tabularnewline{\scriptsize (0.026)}}
\tabularnewline
& & $k$-prototypes
& \shortstack[c]{0.482\tabularnewline{\scriptsize (0.023)}}
& \shortstack[c]{15.525\tabularnewline{\scriptsize (1.896)}}
& \shortstack[c]{0.543\tabularnewline{\scriptsize (0.019)}}
& \shortstack[c]{0.008\tabularnewline{\scriptsize (0.007)}}
\tabularnewline

\midrule
\multicolumn{7}{c}{\textbf{W-H}}
\tabularnewline
\midrule
1.00 & 5.0 & FNJM
& \shortstack[c]{\textbf{0.160}\tabularnewline{\scriptsize (0.078)}}
& \shortstack[c]{\textbf{0.280}\tabularnewline{\scriptsize (0.113)}}
& \shortstack[c]{\textbf{0.924}\tabularnewline{\scriptsize (0.109)}}
& \shortstack[c]{\textbf{0.725}\tabularnewline{\scriptsize (0.213)}}
\tabularnewline
& & Fuzzy JM
& \shortstack[c]{0.206\tabularnewline{\scriptsize (0.086)}}
& \shortstack[c]{0.333\tabularnewline{\scriptsize (0.138)}}
& \shortstack[c]{0.895\tabularnewline{\scriptsize (0.113)}}
& \shortstack[c]{0.633\tabularnewline{\scriptsize (0.217)}}
\tabularnewline
& & G-HMRF
& \shortstack[c]{0.330\tabularnewline{\scriptsize (0.054)}}
& \shortstack[c]{1.552\tabularnewline{\scriptsize (0.390)}}
& \shortstack[c]{0.844\tabularnewline{\scriptsize (0.035)}}
& \shortstack[c]{0.453\tabularnewline{\scriptsize (0.133)}}
\tabularnewline
& & JM
& \shortstack[c]{0.210\tabularnewline{\scriptsize (0.097)}}
& \shortstack[c]{1.124\tabularnewline{\scriptsize (1.058)}}
& \shortstack[c]{0.892\tabularnewline{\scriptsize (0.109)}}
& \shortstack[c]{0.629\tabularnewline{\scriptsize (0.210)}}
\tabularnewline
& & $k$-prototypes
& \shortstack[c]{0.311\tabularnewline{\scriptsize (0.083)}}
& \shortstack[c]{4.577\tabularnewline{\scriptsize (1.266)}}
& \shortstack[c]{0.860\tabularnewline{\scriptsize (0.094)}}
& \shortstack[c]{0.534\tabularnewline{\scriptsize (0.187)}}
\tabularnewline

\bottomrule
\end{tabular*}
\end{table}
Figure~\ref{fig:model-performance} provides a graphical comparison of the competing methods across the four simulation scenarios, confirming the superior performance of FNJM.
%

\begin{figure}
    \centering
    \includegraphics[width=.8\linewidth]{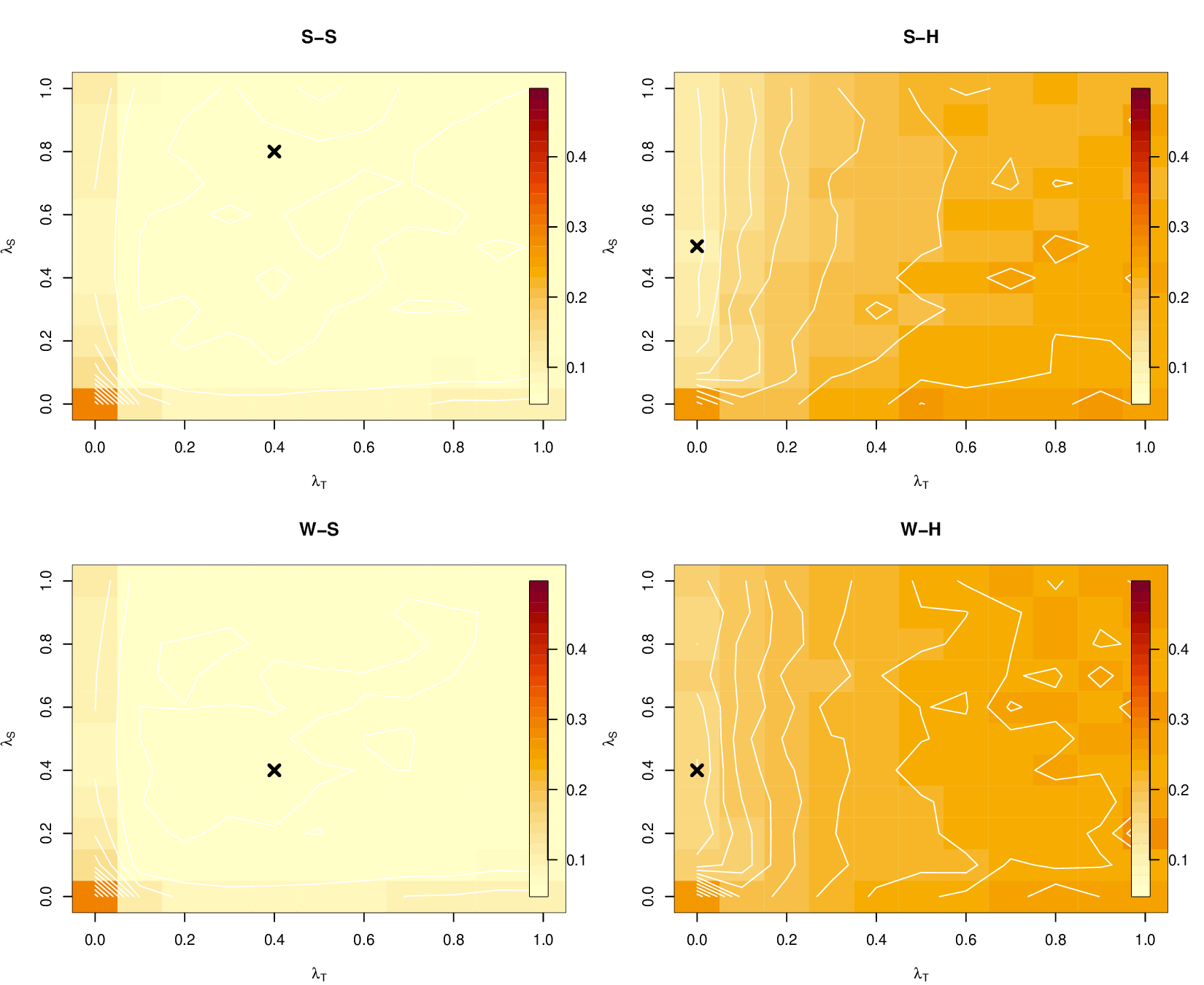}
    \caption{Membership root mean squared error (RMSE) over the temporal and spatial regularization parameters across the four simulation scenarios. The cross identifies the parameter combination minimizing the RMSE.}
    \label{fig:lambdas_heatmap}
\end{figure}

We additionally assess the sensitivity of the model to the temporal and spatial regularization parameters, $\lambda_T$ and $\lambda_S$. Figure~\ref{fig:lambdas_heatmap} reports the membership RMSE over their joint parameter grid. In the soft-membership scenarios (top-left and bottom-left panels), most non-zero combinations yield near-optimal performance, indicating substantial robustness to the choice of the regularization parameters. As expected, larger values of $\lambda_S$ are preferred under stronger spatial dependence. In the nearly hard-membership scenarios, the choice of $\lambda_S$ becomes more influential, particularly under weak spatial dependence, where the graph provides less information for recovering the latent memberships.

\begin{figure}
    \centering
    \includegraphics[width=.8\linewidth]{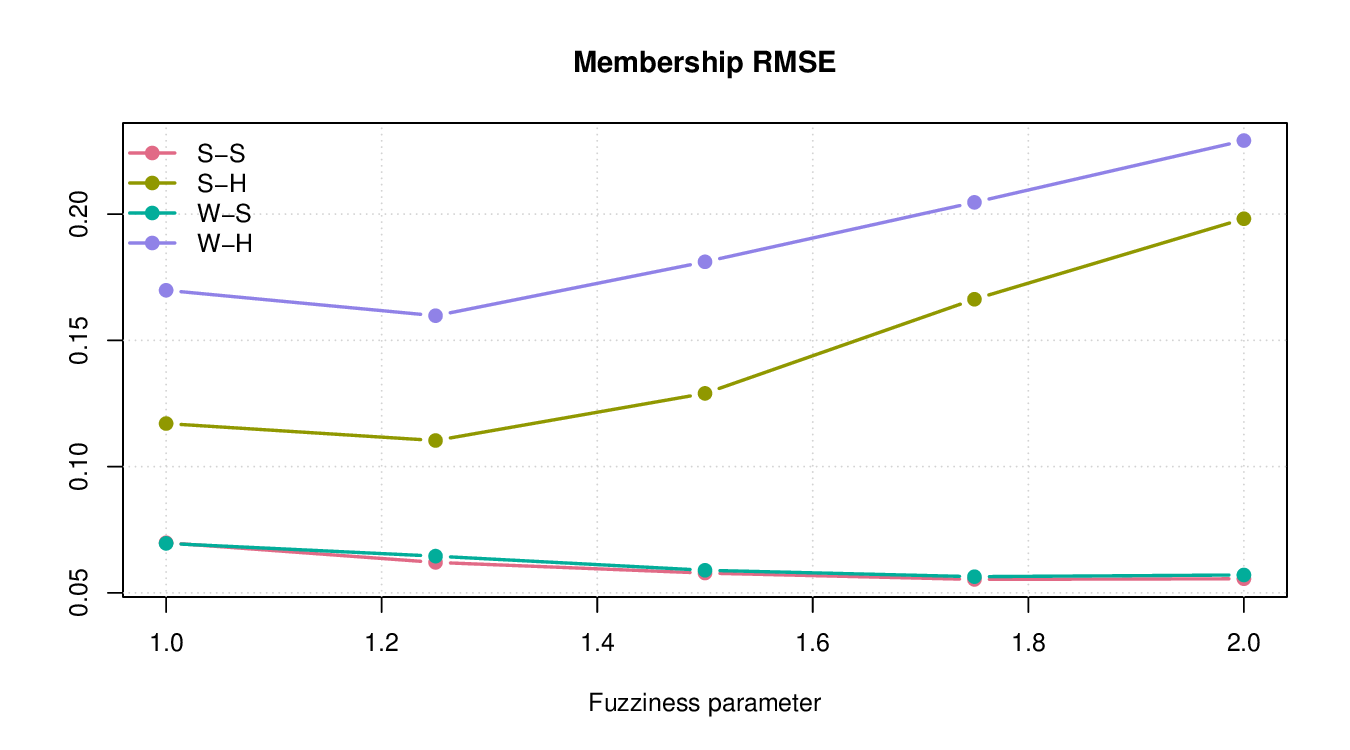}
    \caption{Membership root mean squared error (RMSE) across values of the fuzziness parameter for the four simulation scenarios.}
    \label{fig:fuzz_prof}
\end{figure}

Figure~\ref{fig:fuzz_prof} reports the membership RMSE across values of the fuzziness parameter for each simulation scenario. In the soft-membership scenarios, performance is largely insensitive to this choice. By contrast, in the nearly hard-membership scenarios, lower values are preferred, with the best performance generally obtained for values below approximately \(1.25\).

\section{Application to traffic data}
\label{sec:app}


We apply the proposal to data from the California Department of Transportation
Performance Measurement System (Caltrans PeMS), a widely used benchmark data source for
spatio-temporal traffic analysis \citep{guo2019attention}. In particular, we consider the dataset referred to as PEMS04, comprising measurements from \(M=307\)
monitoring stations in Caltrans District~4. The original data are available, subject to free registration, 
from
\url{https://pems.dot.ca.gov/?dnode=Clearinghouse&type=station_5min&district_id=4&submit=Submit}.

At each station and time point, three continuous traffic variables are recorded: traffic flow (the number of vehicles passing the sensor during a fixed time interval), road occupancy (the proportion of time the sensor is occupied, ranging from 0 to 1), and average vehicle speed in mph.
The original observations are recorded every five minutes. We consider hourly averages of the
first seven days,
obtaining \(T=168\) observations for each station. The resulting dataset comprises \(168 \times 307 = 51{\,}576\) space-time observations.

In order to display the
estimated clusters over the road network, we match station identifiers  with the corresponding geographic metadata,
available from
\url{https://pems.dot.ca.gov/?dnode=Clearinghouse&type=meta&district_id=4&submit=Submit}. The spatial graph is constructed
from the available station connections and inter-station distances. For each
connected pair, the raw spatial weight is defined through a Gaussian kernel,
with bandwidth equal to the median observed edge distance; the resulting
weights are subsequently symmetrically degree-normalized.

\begin{figure}
    \centering
    \includegraphics[width=\linewidth]{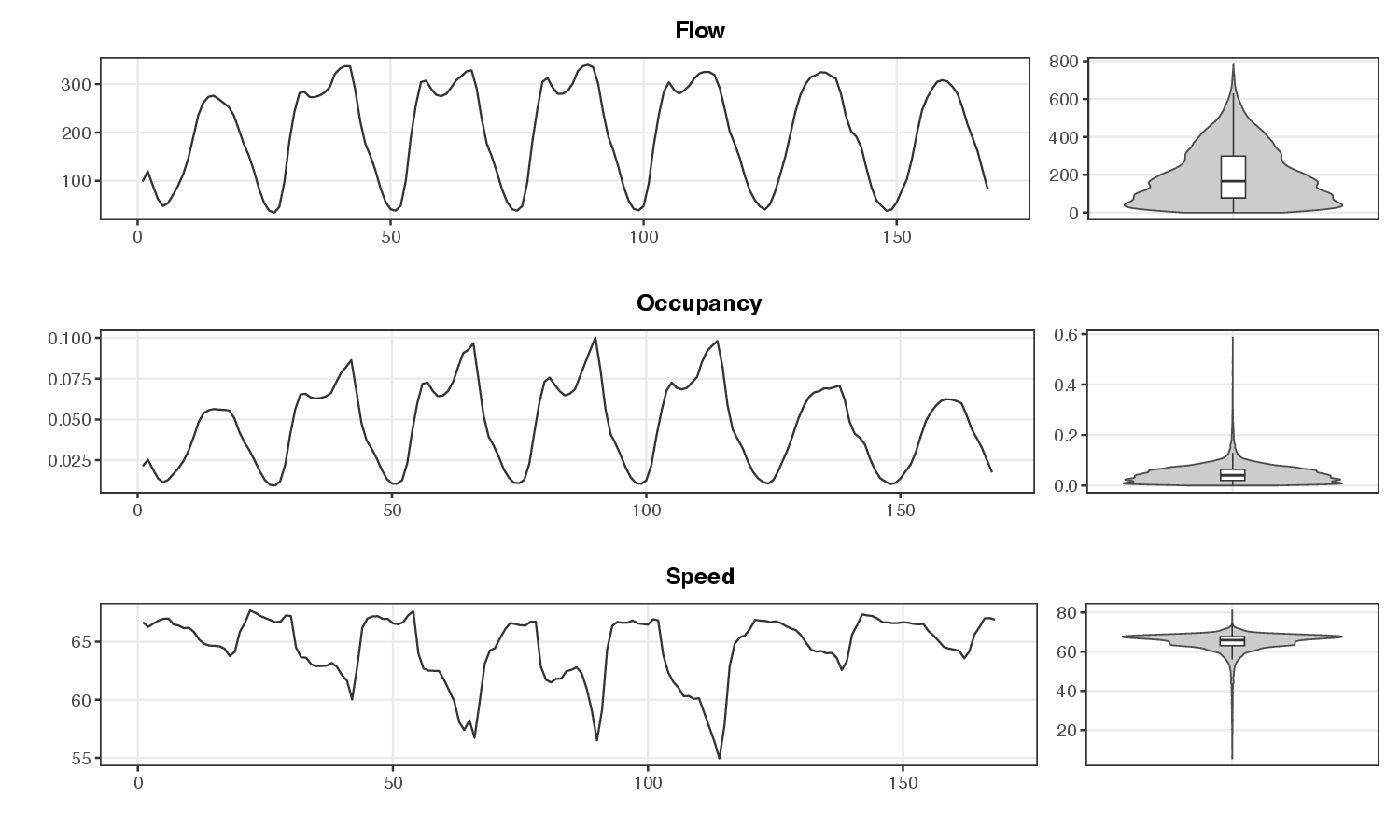}
    \caption{Time series and violin plots of traffic flow, occupancy, and speed.
}
    \label{fig:descr_plots}
\end{figure}

Figure~\ref{fig:descr_plots} presents the time series (averaged across all traffic sensors) and violin plots of traffic flow, occupancy, and speed. The plots reveal a clear day-night cycle, with flow and occupancy increasing during daytime hours and declining overnight. The violin plots indicate right-skewed distributions for flow and occupancy and a pronounced left tail for speed. The unconditional medians are \(166.42\) for flow, \(0.0404\) for occupancy, and \(65.90\) mph for speed, with corresponding standard deviations of \(147.97\), \(0.0365\), and \(6.00\), respectively. Flow and occupancy are strongly positively correlated \((\rho=0.710)\), whereas speed is negatively correlated with both flow \((\rho=-0.329)\) and, more markedly, occupancy \((\rho=-0.766)\).

\subsection{Results}

We select the number of clusters $K$, the fuzziness parameter $\phi$, and the regularization parameters $\lambda_T$ and $\lambda_S$, through
a grid search over
\[
K\in\{2,3,4\},\qquad
\phi\in\{1,1.05,\ldots,1.50\}, \qquad \lambda_T,\lambda_S\in\{0,0.1,\ldots,1\}.
\]
For each candidate model, we compute the fuzzy silhouette index of
\citet{campello2006fuzzy} separately at each time point using
Gower dissimilarities, and use its median over the \(168\) time points as
the selection criterion.
The best-performing configuration is
\(
K=2,\,
\phi=1.15,\,
\lambda_T=1,\,
\lambda_S=0.9,
\)
with a median fuzzy silhouette of \(0.875\). 

The estimated cluster prototypes are reported in Table~\ref{tbl:prototypes}.
Cluster~1 is characterized by substantially higher flow and occupancy, with estimated prototype values of \(312.50\) and \(0.0619\), respectively, together with a lower speed of \(64.23\) mph. We therefore interpret it as a \textit{congested} or high-traffic state. Conversely, Cluster~2, with lower flow (\(99.75\)) and occupancy (\(0.0258\)) and a higher speed (\(66.76\) mph), represents the \textit{uncongested}, low-traffic state.
\begin{table}
\caption{Estimated prototypes for the two latent traffic states.}
\label{tbl:prototypes}
\begin{tabular*}{\textwidth}{@{\extracolsep{\fill}}LCCC@{}}
\toprule
 & Flow & Occupancy & Speed (mph) \\
\midrule
Cluster 1 & 312.50 & 0.0619 & 64.23 \\
Cluster 2 &  99.75 & 0.0258 & 66.76 \\
\bottomrule
\end{tabular*}
\end{table}

\begin{figure}
    \centering
    \includegraphics[width=0.95\textwidth]{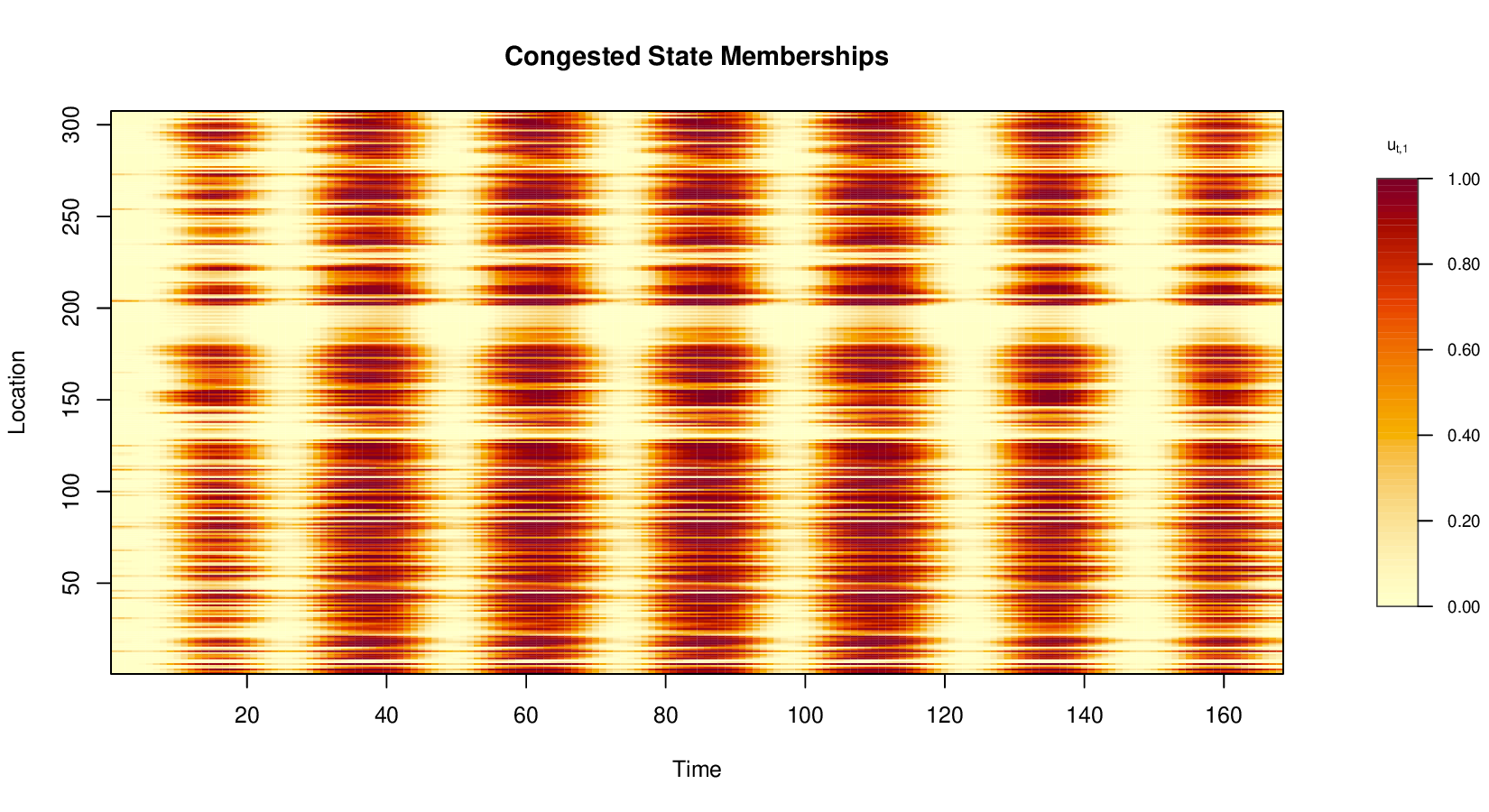}
    \caption{Estimated membership in the congested state (Cluster~1) across
    the \(168\) hourly observations and \(307\) monitoring stations. Darker
    colours indicate a larger estimated membership in the congested state.}
    \label{fig:congested-membership}
\end{figure}

Figure~\ref{fig:congested-membership} displays the estimated membership in the congested state for each station and time point, highlighting both persistent congestion patterns and observations with more uncertain state assignments. It reveals a clear daily cycle, with congestion typically increasing during daytime hours, as well as road segments that are systematically less prone to congestion, usually those located outside the city centre.

\begin{figure}
    \centering
    \includegraphics[width=\textwidth]{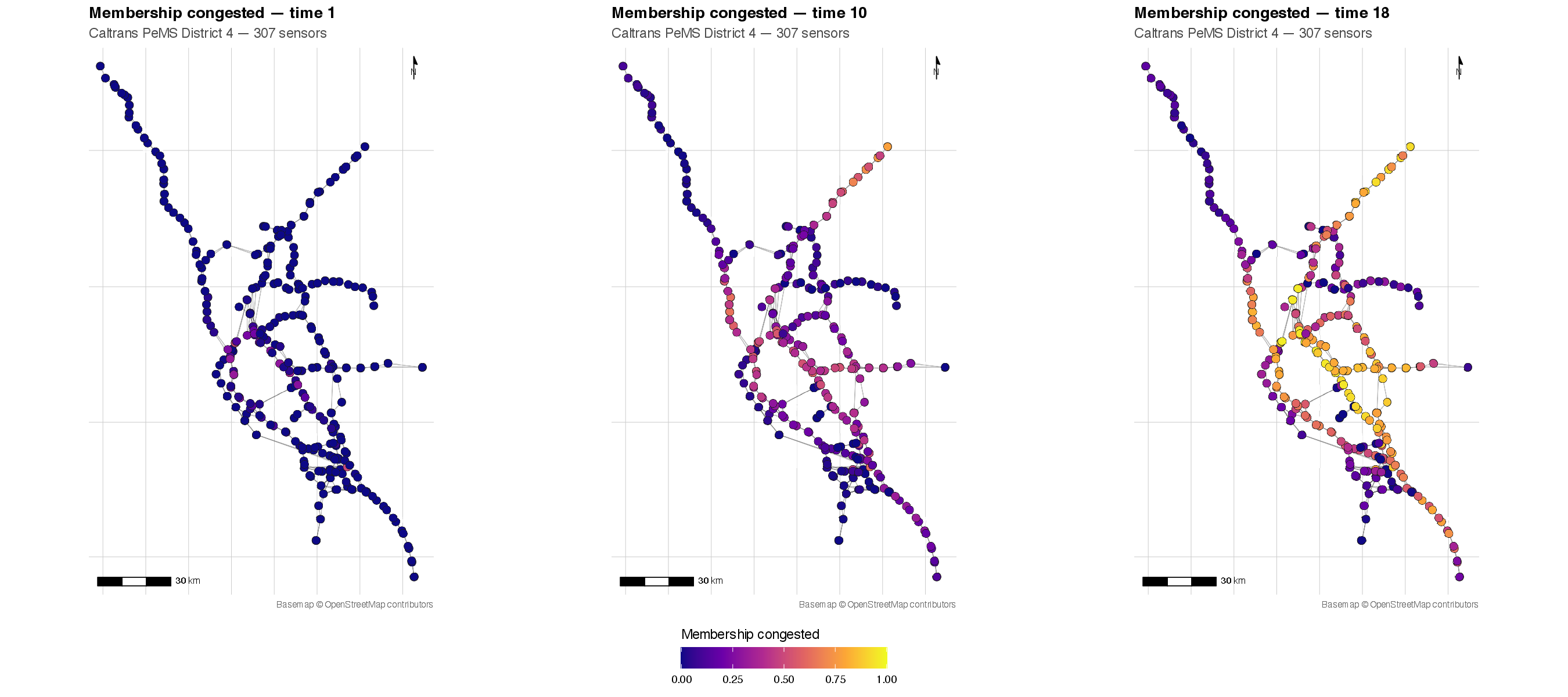}
    \caption{Estimated membership in the congested state at time indices
\(t=1\), \(t=10\), and \(t=18\), shown from left to right. Node colours
represent the estimated membership \(u_{tm1}\),$m=1,\ldots,M$, with larger values indicating
stronger evidence of congestion; lines represent connections in the spatial
graph.}
    \label{fig:traffic-snapshots}
\end{figure}

Figure~\ref{fig:traffic-snapshots} shows three snapshots of the spatial membership in the congested state at different times of day. The panels reveal low congestion during nighttime hours, increased activity in the late morning, and widespread congestion in the late afternoon, associated with the evening commute. They also illustrate how traffic regimes evolve across the road network while remaining spatially coherent among connected monitoring stations.
An interactive visualization of the estimated dynamic traffic clusters is
available at \url{https://fpcortese.shinyapps.io/dynamic-traffic-clustering/}. 

\section{Discussion}
\label{sec:disc}

We introduced a fuzzy network jump model for the dynamic clustering of graph-structured data. The framework combines temporal and spatial regularization to promote persistent and spatially coherent memberships while retaining the ability to detect regime changes. Its flexibility stems from the general representation of the graph, whose edge weights may encode adjacency, distance, connectivity strength, or other application-specific relationships.

The simulation study shows that the proposed method accurately recovers latent memberships across different levels of spatial dependence and cluster overlap. The application to traffic-network data further demonstrates its ability to identify interpretable traffic regimes and describe their evolution over time and across connected road segments.
A limitation of the empirical application is that the traffic network is represented as an undirected graph, so the model cannot distinguish between traffic flows occurring in opposite directions along the same connection.

Future work may consider robust extensions for handling outliers and missing data. A promising direction is to incorporate the cellwise contamination paradigm of \citet{zaccaria2026robust}, allowing anomalous entries to be identified and downweighted without discarding entire observations.

\appendix

\section{Additional simulation results}
\label{app:add_simstud}

In this section, we replicate the simulation study of Section~\ref{sec:simstud} using a substantially denser estimation graph. Specifically, we fix the distance threshold at \(h=0.28\) for all scenarios and replications, yielding a graph density of approximately \(20\%\), which represents a more challenging setting. 


Table~\ref{tbl:add_simulation-results} reports the median RMSE, CE, BAC, and ARI over 100 replicates for each method and scenario, with standard deviations in parentheses. As before, FNJM achieves the best performance across all scenarios, attaining the lowest RMSE and CE and the highest BAC and ARI. Its advantage is particularly pronounced under strong spatial dependence, where it reaches a BAC of \(0.978\) and an ARI of \(0.916\) in the S-H scenario. Among the competing methods, Fuzzy JM generally performs best, while the relative performance of JM, G-HMRF, and \(k\)-prototypes varies across scenarios and metrics. The W-S scenario is the most challenging in terms of classification performance, with FNJM attaining a BAC of \(0.727\) and an ARI of \(0.236\). This is expected because weak spatial dependence provides limited information from the graph, while soft memberships make the latent states less clearly separated.

\begin{table}
\caption{Simulation results for each method across scenarios under the denser graph specification. Entries report median performance over 100 replicates, with standard deviations in parentheses. For each scenario, the lowest median RMSE and CE and the highest median BAC and ARI are shown in bold.}
\label{tbl:add_simulation-results}
\begin{tabular*}{\textwidth}{@{\extracolsep{\fill}}CCLCCCC@{}}
\midrule
$\alpha$ & $\tau$ & Method & RMSE & CE & BAC & ARI
\tabularnewline
\midrule

\multicolumn{7}{c}{\textbf{S-S}}
\tabularnewline
\midrule
0.01 & 0.2 & FNJM
& \shortstack[c]{\textbf{0.036}\tabularnewline{\scriptsize (0.073)}}
& \shortstack[c]{\textbf{0.687}\tabularnewline{\scriptsize (0.059)}}
& \shortstack[c]{\textbf{0.805}\tabularnewline{\scriptsize (0.146)}}
& \shortstack[c]{\textbf{0.402}\tabularnewline{\scriptsize (0.245)}}
\tabularnewline
& & Fuzzy JM
& \shortstack[c]{0.087\tabularnewline{\scriptsize (0.007)}}
& \shortstack[c]{0.700\tabularnewline{\scriptsize (0.003)}}
& \shortstack[c]{0.582\tabularnewline{\scriptsize (0.038)}}
& \shortstack[c]{0.028\tabularnewline{\scriptsize (0.022)}}
\tabularnewline
& & G-HMRF
& \shortstack[c]{0.459\tabularnewline{\scriptsize (0.021)}}
& \shortstack[c]{3.752\tabularnewline{\scriptsize (0.673)}}
& \shortstack[c]{0.548\tabularnewline{\scriptsize (0.034)}}
& \shortstack[c]{0.009\tabularnewline{\scriptsize (0.009)}}
\tabularnewline
& & JM
& \shortstack[c]{0.168\tabularnewline{\scriptsize (0.099)}}
& \shortstack[c]{0.766\tabularnewline{\scriptsize (1.546)}}
& \shortstack[c]{0.575\tabularnewline{\scriptsize (0.050)}}
& \shortstack[c]{0.027\tabularnewline{\scriptsize (0.031)}}
\tabularnewline
& & $k$-prototypes
& \shortstack[c]{0.490\tabularnewline{\scriptsize (0.019)}}
& \shortstack[c]{16.183\tabularnewline{\scriptsize (1.542)}}
& \shortstack[c]{0.544\tabularnewline{\scriptsize (0.017)}}
& \shortstack[c]{0.009\tabularnewline{\scriptsize (0.007)}}
\tabularnewline

\midrule
\multicolumn{7}{c}{\textbf{S-H}}
\tabularnewline
\midrule
0.01 & 5.0 & FNJM
& \shortstack[c]{\textbf{0.055}\tabularnewline{\scriptsize (0.101)}}
& \shortstack[c]{\textbf{0.209}\tabularnewline{\scriptsize (0.121)}}
& \shortstack[c]{\textbf{0.978}\tabularnewline{\scriptsize (0.107)}}
& \shortstack[c]{\textbf{0.916}\tabularnewline{\scriptsize (0.235)}}
\tabularnewline
& & Fuzzy JM
& \shortstack[c]{0.209\tabularnewline{\scriptsize (0.081)}}
& \shortstack[c]{0.345\tabularnewline{\scriptsize (0.128)}}
& \shortstack[c]{0.882\tabularnewline{\scriptsize (0.124)}}
& \shortstack[c]{0.606\tabularnewline{\scriptsize (0.240)}}
\tabularnewline
& & G-HMRF
& \shortstack[c]{0.301\tabularnewline{\scriptsize (0.048)}}
& \shortstack[c]{1.526\tabularnewline{\scriptsize (0.388)}}
& \shortstack[c]{0.880\tabularnewline{\scriptsize (0.028)}}
& \shortstack[c]{0.533\tabularnewline{\scriptsize (0.118)}}
\tabularnewline
& & JM
& \shortstack[c]{0.228\tabularnewline{\scriptsize (0.123)}}
& \shortstack[c]{1.270\tabularnewline{\scriptsize (2.634)}}
& \shortstack[c]{0.866\tabularnewline{\scriptsize (0.134)}}
& \shortstack[c]{0.574\tabularnewline{\scriptsize (0.258)}}
\tabularnewline
& & $k$-prototypes
& \shortstack[c]{0.323\tabularnewline{\scriptsize (0.093)}}
& \shortstack[c]{4.633\tabularnewline{\scriptsize (2.317)}}
& \shortstack[c]{0.853\tabularnewline{\scriptsize (0.122)}}
& \shortstack[c]{0.500\tabularnewline{\scriptsize (0.203)}}
\tabularnewline

\midrule
\multicolumn{7}{c}{\textbf{W-S}}
\tabularnewline
\midrule
1.00 & 0.2 & FNJM
& \shortstack[c]{\textbf{0.049}\tabularnewline{\scriptsize (0.057)}}
& \shortstack[c]{\textbf{0.689}\tabularnewline{\scriptsize (0.045)}}
& \shortstack[c]{\textbf{0.727}\tabularnewline{\scriptsize (0.099)}}
& \shortstack[c]{\textbf{0.236}\tabularnewline{\scriptsize (0.138)}}
\tabularnewline
& & Fuzzy JM
& \shortstack[c]{0.088\tabularnewline{\scriptsize (0.011)}}
& \shortstack[c]{0.699\tabularnewline{\scriptsize (0.005)}}
& \shortstack[c]{0.588\tabularnewline{\scriptsize (0.032)}}
& \shortstack[c]{0.033\tabularnewline{\scriptsize (0.017)}}
\tabularnewline
& & G-HMRF
& \shortstack[c]{0.439\tabularnewline{\scriptsize (0.021)}}
& \shortstack[c]{4.820\tabularnewline{\scriptsize (0.688)}}
& \shortstack[c]{0.548\tabularnewline{\scriptsize (0.023)}}
& \shortstack[c]{0.010\tabularnewline{\scriptsize (0.007)}}
\tabularnewline
& & JM
& \shortstack[c]{0.172\tabularnewline{\scriptsize (0.076)}}
& \shortstack[c]{0.771\tabularnewline{\scriptsize (0.811)}}
& \shortstack[c]{0.592\tabularnewline{\scriptsize (0.042)}}
& \shortstack[c]{0.036\tabularnewline{\scriptsize (0.026)}}
\tabularnewline
& & $k$-prototypes
& \shortstack[c]{0.482\tabularnewline{\scriptsize (0.023)}}
& \shortstack[c]{15.525\tabularnewline{\scriptsize (1.896)}}
& \shortstack[c]{0.543\tabularnewline{\scriptsize (0.019)}}
& \shortstack[c]{0.008\tabularnewline{\scriptsize (0.007)}}
\tabularnewline

\midrule
\multicolumn{7}{c}{\textbf{W-H}}
\tabularnewline
\midrule
1.00 & 5.0 & FNJM
& \shortstack[c]{\textbf{0.182}\tabularnewline{\scriptsize (0.061)}}
& \shortstack[c]{\textbf{0.301}\tabularnewline{\scriptsize (0.089)}}
& \shortstack[c]{\textbf{0.913}\tabularnewline{\scriptsize (0.081)}}
& \shortstack[c]{\textbf{0.691}\tabularnewline{\scriptsize (0.167)}}
\tabularnewline
& & Fuzzy JM
& \shortstack[c]{0.206\tabularnewline{\scriptsize (0.086)}}
& \shortstack[c]{0.333\tabularnewline{\scriptsize (0.138)}}
& \shortstack[c]{0.895\tabularnewline{\scriptsize (0.113)}}
& \shortstack[c]{0.633\tabularnewline{\scriptsize (0.217)}}
\tabularnewline
& & G-HMRF
& \shortstack[c]{0.313\tabularnewline{\scriptsize (0.045)}}
& \shortstack[c]{1.506\tabularnewline{\scriptsize (0.369)}}
& \shortstack[c]{0.860\tabularnewline{\scriptsize (0.028)}}
& \shortstack[c]{0.494\tabularnewline{\scriptsize (0.113)}}
\tabularnewline
& & JM
& \shortstack[c]{0.210\tabularnewline{\scriptsize (0.097)}}
& \shortstack[c]{1.124\tabularnewline{\scriptsize (1.058)}}
& \shortstack[c]{0.892\tabularnewline{\scriptsize (0.109)}}
& \shortstack[c]{0.629\tabularnewline{\scriptsize (0.210)}}
\tabularnewline
& & $k$-prototypes
& \shortstack[c]{0.311\tabularnewline{\scriptsize (0.083)}}
& \shortstack[c]{4.577\tabularnewline{\scriptsize (1.266)}}
& \shortstack[c]{0.860\tabularnewline{\scriptsize (0.094)}}
& \shortstack[c]{0.534\tabularnewline{\scriptsize (0.187)}}
\tabularnewline

\bottomrule
\end{tabular*}
\end{table}



\clearpage
\bibliographystyle{cas-model2-names}

\bibliography{cas-refs}



\end{document}